\documentclass[aps,pre,twocolumn,superscriptaddress,amsmath,floatfix,amssymb]{revtex4}
\usepackage{psfig}
\usepackage{graphics}
\usepackage{graphicx}
\usepackage{epsfig}
\usepackage{psfig}
\usepackage{amsmath}

\begin{document}
\title{Dynamical Exchanges in Facilitated Models of Supercooled Liquids}

\author{YounJoon Jung}
\affiliation{Department of Chemistry, University of California, 
Berkeley, CA 94720-1460}

\author{Juan P. Garrahan}
\affiliation{School of Physics and Astronomy, University of 
Nottingham, Nottingham, NG7 2RD, UK}

\author{David Chandler}
\affiliation{Department of Chemistry, University of California, 
Berkeley, CA 94720-1460}

\date{\today}

\begin{abstract}
We investigate statistics of dynamical exchange events in
coarse--grained models of supercooled liquids in spatial dimensions
$d=1$, 2, and 3. The models, based upon the concept of dynamical
facilitation, capture generic features of statistics of exchange times
and persistence times.  Here, distributions for both times are
related, and calculated for cases of strong and fragile glass formers
over a range of temperatures. Exchange time distributions are shown to
be particularly sensitive to the model parameters and dimensions, and
exhibit more structured and richer behavior than persistence time
distributions.  Mean exchange times are shown to be Arrhenius,
regardless of models and spatial dimensions.  Specifically, $\langle
t_{\rm x}\rangle \sim c^{-2}$, with $c$ being the excitation
concentration.  Different dynamical exchange processes are identified
and characterized from the underlying trajectories.  We discuss
experimental possibilities to test some of our theoretical findings.
\end{abstract}


\maketitle

\section{Introduction}
\label{sec1}

Dynamical arrest of liquids as they approach the glass transition is a
topic of much current research. 
\cite{ediger-jpc-96,angell-sci-95,debenedetti-nature-01,kob-rev-03}  This
arrest entails such features as non--exponential relaxation and
precipitous non--Arrhenius temperature dependence of transport
properties.  At its heart lies the notion of dynamic heterogeneity. 
\cite{DHexp,DHnum, vandenbout,vidalrussell-nature-00,sillescu-jncs-99,ediger-arpc-00,glotzer-jncs-00,richert-jpcm-02}
Namely, as the temperature decreases towards the glass transition, mobility
develops spatial inhomogeneity, and as time progresses local mobility
undergoes dynamical changes.  We have interpreted these phenomena
\cite{garrahan-prl-02,garrahan-pnas-03,berthier-pre-03,jung-pre-04,berthier-epl-05}
in terms of local excitations of mobility in space that propagate in
time with facilitated dynamics. 
\cite{fredrickson-prl-84,palmer-prl-84,jackle-zphysb-91,butler-jcp-91}
The excitations thus form lines in space--time.  Dynamic scaling is
described naturally in terms of the statistics of these lines. 
\cite{garrahan-prl-02,garrahan-pnas-03,wbg-04,berthier-pre-04}

The work we present in this paper is motivated by recent experiments
that directly detect local fluctuations of dynamics in supercooled
liquids. 
\cite{ediger-arpc-00,vandenbout, bartko-prl-02,vidalrussell-nature-00,richert-jpcm-02}
For example, single molecule rotational experiments
\cite{vandenbout}
show that changes in local
dynamics are similar to random telegraph noise, where local spatial
regions exhibit dynamical exchanges between fast and slow
dynamics. Similar stochastic behavior is also observed in local
dielectric fluctuation experiments. \cite{vidalrussell-nature-00}  In
those experiments a local spatial region exhibits dynamical exchanges
between fast and slow dynamics, thus yielding direct confirmation of
dynamic heterogeneity.

Considering these experiments, we establish generic features of
statistical properties of dynamical exchange events and show how these
features manifest themselves in experimental observables. In the low
temperature regime, dynamics in supercooled liquids is dominated by
fluctuations, not the mean.  Therefore, it is pertinent to study the
whole distribution in order to gain insights on the microscopic nature
of dynamics in supercooled liquids and glasses.

We do so in this paper in the context of coarse--grained facilitated
models. In Sec.~\ref{sec2}, we define these models. In
Sec.~\ref{sec3}, we define exchange and persistence times, and derive
an analytical relationship between the distributions of exchange and
persistence times. In Sec.~\ref{sec4}, qualitative features of the
exchange time distribution are presented. Numerical results of
distributions of exchange and persistence times for the models are
described in Sec.~\ref{sec5}.  In Sec.~\ref{sec6} we present results
of numerical simulations of dynamic bleaching experiments. We conclude
in Sec.~\ref{sec7} with discussions of possible experimental
measurements aimed at testing our theoretical predictions.

\section{Models of Glass Formers}
\label{sec2}

We assume that a kinetically constrained model
\cite{garrahan-prl-02,garrahan-pnas-03,fredrickson-prl-84,jackle-zphysb-91,ritort-advphys-03}
is obtained through coarse graining over a microscopic time scale
$\delta t$ (e.g., larger than the molecular vibrational time scale),
and also over a microscopic length scale $\delta l$ (e.g., larger than
the equilibrium correlation length).  The dimensionless Hamiltonian
for the model is,
\begin{equation} {\cal H}=\sum_{i=1}^{N} n_{i}, \ \ (n_i=0,
  1). \label{hamil}
\end{equation}  
Here, $n_i$ refers to a state of lattice site $i$ at ${\bf x}_i$,
where $n_i=1$ coincides with lattice site being a spatially unjammed
region (i.~e., carrying mobility), while $n_{i}=0$ coincides with it
being a jammed region (i.~e., not carrying mobility). We thus call
$n_i$ the ``mobility field''. In spatial dimensions $d=1$, 2, and 3,
the lattice is linear, square, and cubic, respectively.  The number of
sites, $N$, specifies the size of the system. From Eq.~(\ref{hamil}),
thermodynamics is trivial, and the equilibrium concentration of
defects or excitations is given by
\begin{equation} 
c=\langle n_i\rangle = \frac{1}{1 + \exp (1/T)},
\end{equation}  
where $T$ is a reduced temperature.

The dynamics of these models obeys detailed balance with local
dynamical rules that depend on the configuration of the lattice site
$i$ as well as those of its neighbors. Namely,
\begin{eqnarray}
n_{i}(t)=0 \xrightarrow{k_{i}^{(+)}} n_{i}(t+\delta t)= 1, \\
n_{i}(t)=1 \xrightarrow{k_{i}^{(-)}} n_{i}(t+\delta t)= 0,
\end{eqnarray}
where
\begin{eqnarray}
k_{i}^{(+)}&=& e^{-1/T} f_{i}\left(\left\{ n_{\bf x} \right \}\right),
   \label{k+} \\ k_{i}^{(-)}&=& f_{i}\left(\left\{ n_{\bf x} \right
   \}\right). \label{k-}
\end{eqnarray}
The function $f_{i}\left(\left\{ n_{\bf x}
   \right\}\right)=f_{i}(n_{1},n_{2},\cdots,n_{N})$ reflects dynamical
   facilitation.

In the Fredrickson--Andersen (FA) model, \cite{fredrickson-prl-84} a
change of state at site $i$ can occur from $t$ to $t+\delta t$,
$n_i(t+\delta t) = 1- n_i(t)$, if at least one of $2d$ nearest
neighbors of the $i$th site is excited. For example, in $d=1$ case, 
\cite{fredrickson-prl-84} $n_i(t+\delta t) = 1 - n_i(t)$ is allowed
only if $n_j(t)=1$ where $x_j= x_i\pm \delta l$.  In contrast, in the
$d=1$ East model case, \cite{jackle-zphysb-91} one needs an excited
neighbor in specific directions in order to have a state change at
site $i$. For instance, $n_j(t)=1$ such that $x_j=x_i - \delta l$.

In $d=3$ cases, $n_i(t)$ at ${\bf x}_{i}=(x,y,z)$ can make a change if
at least there is one nearest neighbor site $j$ such that $n_j(t)=1$
at ${\bf x}_{j}=(x\pm \delta l, y, z)$ or $(x, y\pm \delta l , z)$ or
$(x, y, z\pm \delta l)$ (FA model) or ${\bf x}_{j}=(x-\delta l, y, z)$
or $(x, y-\delta l, z)$ or $(x, y, z-\delta l)$ (East model). The
facilitation function for the FA and East model, $f_{i}\left( \left\{
n_{\bf x}\right\} \right)$, is therefore given by,
\begin{eqnarray}
  \left[ f_{i}\left(\left\{ n_{\bf x} \right\} \right)\right ]_{\rm
  FA} &=& 1-{\prod_{l=1}^{d}} (1-n_{ {{\bf x}_{i}}+{\delta l {\hat{\bf
  u}}_l}}) (1-n_{ {{\bf x}_{i}}-{\delta l {\hat{\bf u}}_l}}),
  \nonumber \\
\label{fifa} \\
\left[ f_{i}\left(\left\{ n_{\bf x}\right\}\right)\right ]_{\rm East}
  &=& 1-{\prod_{l=1}^{d}} (1-n_{ {{\bf x}_{i}}-{\delta l {\hat{\bf
  u}}_l}}), \label{fieast}
\end{eqnarray}
where ${\hat{\bf u}}_l$ is a unit vector in the $l$th dimension.  In
the classification scheme of kinetically constrained models given in
Ref.~\onlinecite{ritort-advphys-03}, these kinetic constraints correspond to
one--spin facilitation rule in $d$ dimension.

\section{Exchange and Persistence Times}
\label{sec3}

\begin{figure}
\begin{center}
\epsfig{file=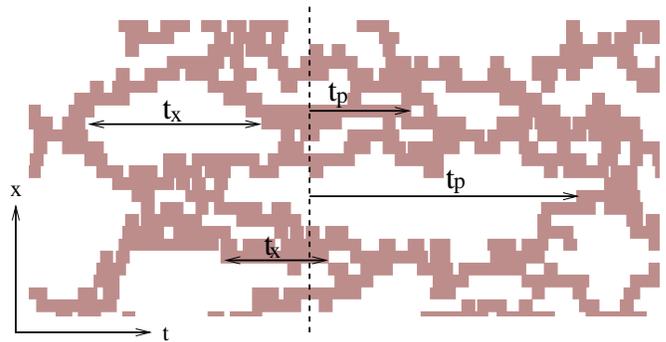, width=\columnwidth}
\caption{Exchange and persistence times are shown in the trajectory of
$d=1$ FA model at $T=0.8$. Shaded regions represents parts of
space-time with excitations, while white regions represent parts with
no excitations.}
\label{traj}
\end{center}
\end{figure}

Relaxation time scales of the glassy systems can be described by a
distribution of {\it persistence times}, $t_{\rm p}$, the time at
which a local region changes its state for the first time once the
trajectory has started at time zero. In the $d=1$ FA model, for
instance, the mean persistence time exhibits Arrhenius behavior at low
temperatures. \cite{berthier-pre-03}

A statistical measure which is useful in studying changes in dynamics
of local environments is the distribution of {\it exchange times},
$t_{\rm x}$, the time duration of given states of mobility fields. For
example, it has been shown that the decoupling of the translational
diffusion from the relaxation times in supercooled liquids near the
glass transition can be explained from the distribution of exchange
times. \cite{jung-pre-04,berthier-epl-05}  Persistence and exchange
times are illustrated in the trajectory of $d=1$ FA model in
Fig.~\ref{traj}.

Exchange and persistence times are different statistical measures of
the same trajectories. There is a general relation between exchange
and persistence time distributions, independent of models. To derive
the relation, we define the variable
\begin{eqnarray}
P_{i}(t;1,\tau)&=&\left[\prod_{t'=0}^{t-\delta t} n_i 
(\tau+t')\right] \left[1-n_i(\tau + t)
\right]. \label{peop}
\end{eqnarray}
It is the density of excitations at time $\tau$ (i.e., $n_{i}=1$) that
persist until time $t+\tau$. A similar expression for the density of
unexcited sites that persist for this time frame, $P_{i}(t;0,\tau)$,
is given by Eq.~(\ref{peop}) with $n_{i}$ changed to $1-n_{i}$, and
$1-n_{i}$ changed to $n_{i}$.

Similarly,
\begin{eqnarray}
X_{i}(t;1, \tau)=&&\frac{1}{\delta t}\left[1-n_i(\tau-\delta t)\right]
\left[\prod_{t'=0}^{t-\delta t} n_i(\tau+t')\right] \nonumber \\
&&\times \left[1-n_i(\tau + t) \right], \label{exop}
\end{eqnarray}
is the density of excitations that exchange over the time interval
between $\tau$ and $t+\tau$. Clearly,
\begin{eqnarray}
X_{i}(t;1,\tau) = \frac{1}{\delta t}\left[ P_{i}(t;1, \tau)
-  P_{i}(t+\delta t; 1,\tau-\delta t)\right ].  \label{ExvsPe}
\end{eqnarray}
Averaging and taking $\delta t \rightarrow 0^{+}$ therefore yields
\begin{equation}
{\cal X}(t;1) = -\frac{\mathrm{d} {\cal P}(t;1)}{\mathrm{d} t},
\label{EvsP}
\end{equation} 
where ${\cal X}(t;1) = \langle X_{i}(t;1,\tau) \rangle$ and ${\cal P}(t;1)=
\langle P_{i}(t;1,\tau) \rangle$.  The normalized probability densities of
exchange and persistence times for an excitation, $p_{\rm x}(t;1)$ and
$p_{\rm p}(t;1)$, respectively, are proportional to ${\cal X}(t;1)$
and ${\cal P}(t;1)$, respectively.  Accounting for normalization
constants therefore gives
\begin{equation}
p_{\rm p}(t;1) = \frac {\int_{t}^{\infty} {\mathrm{d}} t'\, p_{\rm
    x}(t';1)} {\langle t_{\rm x}(1)  \rangle }, \label{PpvsPe1}
\end{equation} 
where $\langle t_{\rm x}(1)\rangle=\int_{0}^{\infty} {\mathrm{d}}t \, t p_{\rm
x}(t;1)$ is the mean exchange time for an excitation.

One can also proceed the same route for persistence and exchange of no
excitation by replacing $n_{i}$ with $1-n_{i}$ in Eqs.~(\ref{peop})
and (\ref{exop}) to find
\begin{equation} 
p_{\rm p}(t;0) = \frac{\int_{t}^{\infty} {\mathrm{d}} t'\, p_{\rm
    x}(t';0)}{\langle t_{\rm x}(0) \rangle }.
\label{PpvsPe0} 
\end{equation} 

The overall exchange and persistence time probability densities,
$p_{\rm x}(t)$, and $p_{\rm p}(t)$, are given by
\begin{eqnarray}
p_{\rm x} (t) &=& \frac{1}{2} \left [ p_{\rm x}(t;0) + p_{\rm
     x}(t;1)\right ], \\
p_{\rm p} (t) &=& (1-c) p_{\rm p}(t;0) + c p_{\rm p}(t;1).
\end{eqnarray}
Using Eqs.~(\ref{PpvsPe1}) and (\ref{PpvsPe0}) with the condition of
detailed balance,
\begin{equation} \frac{\langle t_{\rm x}(1) \rangle }{\langle t_{\rm x}(0)\rangle} =\frac{c}{1-c}, 
\end{equation} \label{detbal}
we find
\begin{equation} p_{\rm p}(t) = \frac{\int_{t}^{\infty} \mathrm{d} t'\, p_{\rm x} (t') }{
{\langle t_{\rm x}\rangle}},
\label{ppepex} \end{equation} 
where $\langle t_{\rm x}\rangle = \frac{1}{2} (\langle t_{\rm x}(0)\rangle + \langle
t_{\rm x}(1)\rangle)$. For a Poisson process, which is valid for
unconstrained dynamics, $p_{\rm x}(t;n) = \tau(n)^{-1}
\exp(-t/\tau(n))$. In that case, the exchange and persistence time
distributions become identical to each other. \cite{renewal}

Moments of the two distributions are related to each other from
Eq.~(\ref{ppepex}) through integration by parts. In particular,
\begin{equation} \langle (t_{\rm p})^m \rangle = \frac{\langle (t_{\rm x})^{m+1}\rangle }{(m+1)!
   \langle t_{\rm x}\rangle}.
\label{mompexppe} \end{equation} 
The moments of the persistence time distributions are always greater
than those of the exchange time distributions when the distributions
are broader than Poissonian.  Correlations between exchange events are
described by correlation functions of the type
\begin{equation}
C_{ij}(t,t'|n,\tau, n',\tau')
= \langle \delta X_{i}(t;n, \tau ) \delta X_{j}(t';n', \tau') \rangle, \label{corr}
\end{equation} 
where $\delta X_{i}(t;n, \tau) \equiv X_{i}(t;n,\tau)- {\cal
   X}(t;n)$.

\section{Analysis of Exchange Time Distributions}
\label{sec4}

We have calculated exchange and persistence time distributions for the
FA and East models by performing Monte Carlo simulations. For the
purpose of numerical efficiency, we have used the continuous time
Monte Carlo algorithm. \cite{bkl-jcompphys-75,newmanbarkema}  To
sample long exchange times (i.e., $\log_{10} t_{\rm x} > 1$), we have
used $N=100 c^{-1}$. To sample short exchange times (i.e., $\log_{10}
t_{\rm x} < 1$), we have used $N=10^{5} c^{-1}$.  The two
distributions obtained were matched at $\log_{10} t_{\rm x}=1$.
Simulations were performed for total times ${\cal T}= 500\ \tau\alpha$, with
$\tau\alpha$ being the relaxation time of the model. Averages were performed
on about 100 independent trajectories in each case.

\begin{figure}
\begin{center}
\epsfig{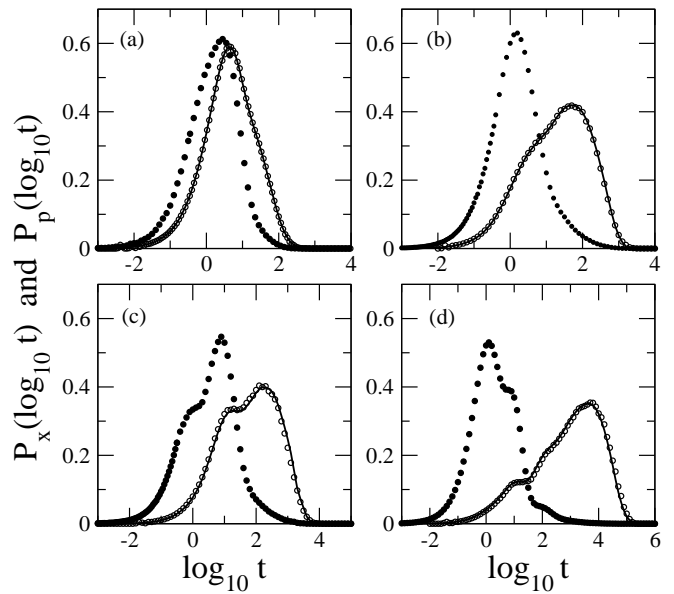}
\caption{Exchange and persistence time distributions for $d=1$ FA and
East models. Here, $P_{\rm x}(\log_{10} t) = t p_{\rm x}(t) \ln(10)$,
and $P_{\rm p}(\log_{10} t)$ is similarly related to $p_{\rm p}(t)$.
Exchange time distribution from simulation (filled circles),
persistence time distribution from simulation (open circles), and
persistence time distribution predicted from Eq.~(\ref{ppepex}) (solid
lines).  (a) $d=1$ FA model at $T=1$, (b) $d=1$ FA model at $T=0.5$,
(c) $d=1$ East model at $T=1$, and (d) $d=1$ East model at $T=0.5$.}
\label{exvspe}
\end{center}
\end{figure}

As an illustration,
we show in Fig.~\ref{exvspe} exchange and persistence time distributions
obtained from numerical simulations, and compare those results with
the prediction of Eq.~(\ref{ppepex}). The comparison verifies
Eq.~(\ref{ppepex}). The peak and principal statistical weight in the 
persistence time distribution occurs at a longer time than that for 
the exchange time distribution.

\begin{figure}
\begin{center}
\epsfig{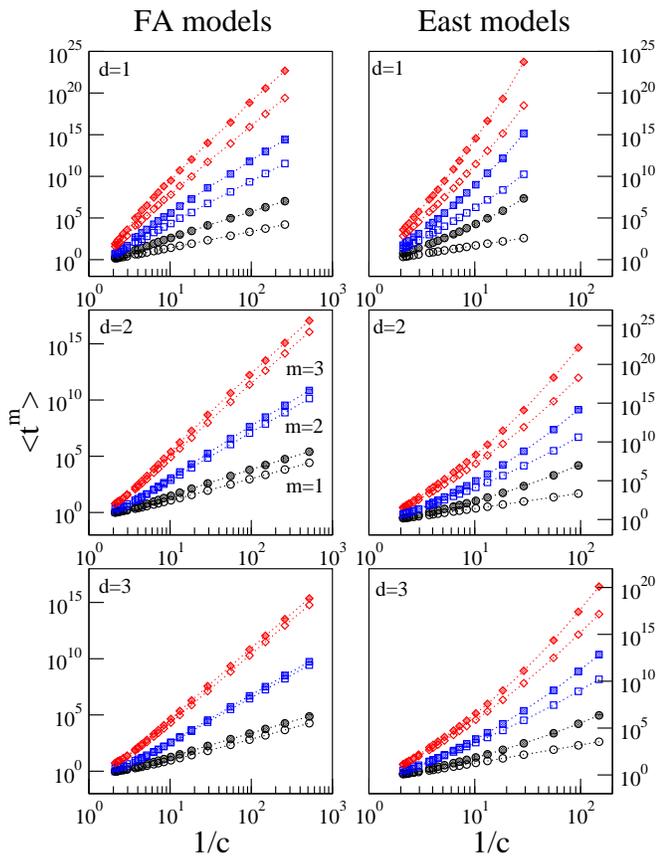}
\caption{The first (circles), second (squares), and third (diamonds)
moments of exchange (open symbols) and persistence (shaded symbols)
time distributions of $d=1$,2, and 3 FA models (left panels) and East
models (right panels) as functions of excitation concentration, $c$.}
\label{tmom}
\end{center}
\end{figure}

\subsection{Moments of Exchange and Persistence Times}

In many experiments, measurements of moments of fluctuating quantities
are more easily made than measurements of the whole
distributions. Studies of moments will yield information on underlying
dynamics. \cite{jung-acp-02,jung-arpc-04}  We have studied temperature
dependence of the moments of exchange and persistence time
distributions, $\langle (t_{\rm x})^m\rangle$ and $\langle (t_{\rm p})^m \rangle$
($m$=1, 2, and 3), respectively, of the FA and East models. Figure
\ref{tmom} shows these moments for dimensionality $d$=1, 2, and 3.

Moments of the exchange and persistence times reveal different
temperature dependence for the FA and the East models. In the FA
models, all the moments of exchange and persistence times increase in
an Arrhenius fashion, $\langle t^{m}\rangle \sim c^{-\alpha_m}$, as the
temperature (or the excitation concentration) decreases. In the East
models only the mean exchange times are Arrhenius. Higher moments of
exchange time distribution and all the moments of persistence time
distribution in the East models are super--Arrhenius.  Further, the
mean persistence time is larger than the mean exchange time in all
cases.  Table \ref{exptable} lists scaling exponents of the moments
for the exchange and persistence times for Arrhenius cases.

\begin{table}
\begin{center}
\begin{tabular}{ccc}
\begin{tabular}{|c|c|c|c|}
\hline
\multicolumn{4}{|c|} {FA models}  \\
\hline\hline
        & $m=1$ & $m=2$ & $m=3$  \\
\hline
$d= 1$ & 2.0(3.2)  &  5.2(6.4)  & 8.4(9.7) \\
\hline
$d= 2$ & 1.9(2.3) & 4.2(4.7) &  6.6(7.0)  \\
\hline
$d= 3$ & 1.9(2.1) & 4.0(4.2) &  6.1(6.3) \\
\hline
\end{tabular} &  &
\begin{tabular}{|c|c|}
\hline
\multicolumn{2}{|c|} {East models}  \\
\hline\hline
        & $m=1$ \\
\hline
$d= 1$ & 2.0  \\
\hline
$d= 2$ & 2.0 \\
\hline
$d= 3$  & 1.9 \\
\hline
\end{tabular}
\end{tabular}
\end{center}
\caption{Scaling exponents for fitting various moments of exchange and
persistence time distributions to Arrhenius form, $\langle t^{m} \rangle \sim
c^{-\alpha_{m}}$ are shown for $d=1$, 2, and 3 FA and East
models. Numbers in the parenthesis are those for persistence time
distributions.  Fitting of the moments to the Arrhenius form was done
for data corresponding to $T\le1$. }
\label{exptable}
\end{table}

\begin{figure}
\begin{center}
\epsfig{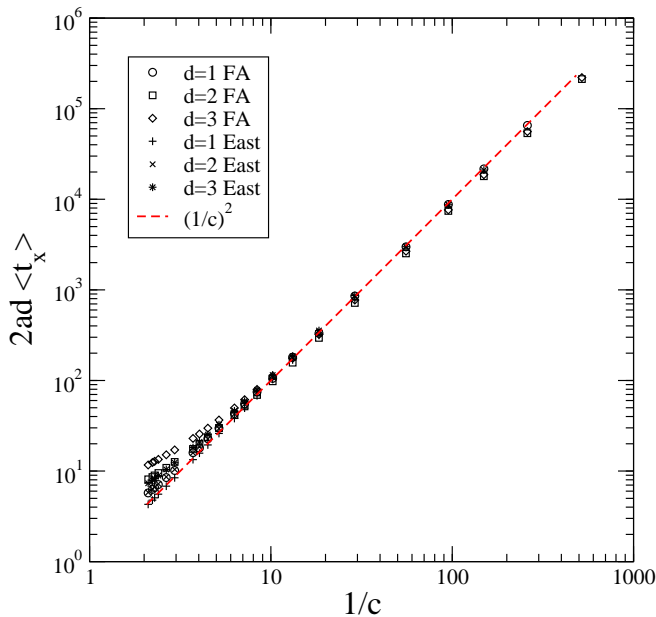}
\caption{Comparison of mean--field approximation, Eq.~(\ref{tmeaneq})
   (dashed line), and simulation results of mean exchange times in
   $d=1$, 2, and 3 FA and East models. $a=2$ for FA models and $a=1$
   for East models.}
\label{tmean}
\end{center}
\end{figure}

We find that the mean exchange time increases approximately as
\begin{equation}
\langle t_{\rm x} \rangle \sim  c^{-2}, \label{c2}
\end{equation} 
in the low temperature limit in all cases, regardless of the models
and dimensionality. To understand this result, we use
Eqs.~(\ref{fifa}) and (\ref{fieast}) to establish that the average
value of facilitation function, $\langle f_i\rangle$, is
\begin{equation}
\langle f_i \rangle = 1 - (1-c)^{a d},
\end{equation} 
where $a=2$ and 1 for FA and East models, respectively. As such, the
mean exchange times for $n=0$ and $n=1$ states obtained from flip
rates in Eqs.~(\ref{k+}) and (\ref{k-}) are
\begin{eqnarray} 
\langle t_{\rm x}(0)\rangle &\approx& 1/\langle k_i^{(+)}\rangle = \exp(1/T) /  \langle f_i \rangle, \\
\langle t_{\rm x}(1)\rangle &\approx& 1/\langle k_i^{(-)}\rangle = 1/ \langle f_i \rangle, \end{eqnarray} 
and the overall mean  exchange time,
$\langle t_{\rm x}\rangle = \frac{1}{2} (\langle t_{\rm x}(0)\rangle + \langle t_{\rm
   x}(1)\rangle )$, is
\begin{equation}
\langle t_{\rm x} \rangle \approx \frac{1}{2 c \langle f_i\rangle} =
{1 \over 2 c [1-(1-c)^{ad}]} \ \rightarrow \ \frac{1}{2 a d c^2}, \
\mbox{as $c\rightarrow 0$}, \label{tmeaneq}
\end{equation} 
which shows the observed scaling of $\langle t_{\rm x}\rangle \sim c^{-2}$ in
the low temperature limit with a proportionality constant that depends
on the dimensionality and model.  Figure \ref{tmean} shows that both
the scaling and the proportionality constant predicted within
Eq.~(\ref{tmeaneq}) agree with the simulation results.

This behavior of the mean exchange time versus the excitation
concentration has at least two experimental implications.  First,
experimental measurements that are insensitive to fluctuations of
exchange times will not distinguish between fragile and strong glass
formers.  In order to distinguish, measurements of higher moments, or
perhaps, the whole distributions will be required. The distribution
could possibly be observed through single molecule experiments. 
\cite{vandenbout,jung-acp-02,jung-arpc-04}  Second, the
universal temperature dependence of the mean exchange time can form a
basis of experimental measurements of excitation concentrations in
real systems.  The specific $c^{-2}$ scaling holds to the extent that
experimental systems can be modeled with single spin-flip facilitation
rules.

In contrast to the behavior of the mean exchange time, the temperature
dependence of mean persistence time, $\langle t_{\rm p}\rangle$, depends on
the model and the dimensionality.  Relaxation times of the models are
given by their mean persistence times, or from Eq.~(\ref{mompexppe}),
the second moment of the exchange times. Temperature dependence of
moments of exchange and persistence times shows that relaxation times
are Arrhenius for the FA models and super-Arrhenius for the East
models.

As the dimensionality of the model increases, differences between the
moments of exchange and persistence times become smaller in the FA
model. The model becomes increasingly mean--field like in higher
dimensions.  No such convergence is found in the East model for all
dimensions investigated.  The East models have weak dimensional
dependence due to the quasi-one dimensional nature of directional
persistence.

\subsection{Stokes--Einstein Breakdown}
For the FA models, which correspond to strong glass formers, the 
translational diffusion constant scales as
\cite{jung-pre-04}
\begin{equation} 
D_{\rm FA}\sim \langle t_{\rm x}\rangle^{-1} \sim c^2.  
\label{diff} 
\end{equation} 
This result is true in all dimensions. It follows from Eq.~(\ref{c2}) 
and the statement that adjacent random walk steps coincide with adjacent exchange events. 
For the East models, however, the result does not hold because 
adjacent exchange events are correlated.  In particular, for the 
hierarchical dynamics of the East models, large bubbles in space-time 
are covered with smaller bubbles, which in turn are covered by yet 
smaller bubbles, and so on. \cite{garrahan-prl-02}   
In that case, higher moments of exchange time distributions exhibit different
temperature dependence from that of the mean.  This behavior is
illustrated in Fig.~\ref{tmom} for the East model, which corresponds
to a fragile liquid.  Correlations between successive exchange events
can be investigated by using correlation function given in
Eq.~(\ref{corr}).  The effects are responsible for non--trivial
scaling relation between the translational diffusion and persistence
times in fragile liquids. \cite{jung-pre-04,jung-unpub,fujara-zphyb-92,chang-jpcb-97,swallen-prl-03,schweizer-jpcb-04}

Indeed, numerical results for the fractional Stokes-Einstein
relationship, 
\begin{equation} 
D \propto \tau_{\alpha}^{-\xi} 
\end{equation} 
give $\xi \approx
2/3, 2/2.3, 2/2.1$ for dimensions $d=1,2,3$ in the case of the FA
models, while $\xi \approx 0.7-0.8$ (very weakly dependent on $d$) in
the case of the East models. \cite{jung-pre-04,jung-unpub}

\subsection{Classifications of Exchange Events}

\begin{figure}
\begin{center}
\epsfig{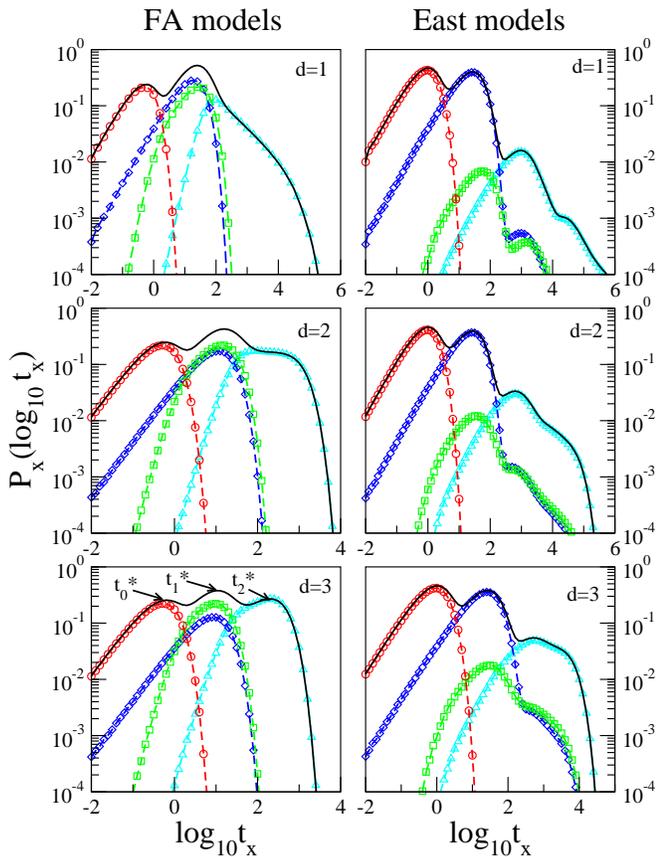}
\caption{Decompositions of exchange time distributions of FA models
(left panels) and East models (right panels) for $d=1$,2, and 3 at
$T=0.3$.  Four sub-distributions -- case 1 (circles), case 2
(squares), case 3 (diamonds), and case 4 (triangles) - add up to the
full distribution (solid line). Positions of three peaks in
$P(\log_{10} t_{\rm x})$ are shown for $d=3$ FA model cases.}
\label{decomp}
\end{center}
\end{figure}

We show in Fig.~\ref{decomp} representative exchange time
distributions.  At short exchange times, $\log_{10}t_{\rm x}<-1$, the
distribution $P_{\rm x}(\log_{10} t_{\rm x})$ follows a power-law
behavior with a slope 1 in all cases, $P_{\rm x}(\log_{10} t_{\rm x})
\sim \log_{10}t_{\rm x}$. This is because $P_{\rm x}(\log_{10} t_{\rm
x}) \sim t_{\rm x} p_{\rm x}(t_{\rm x})$, and Poissonian statistics is
obeyed, $p_{\rm x}(t_{\rm x}) \sim \exp(-t_{\rm x})$, for short times.

Multiple peak structures develop in the exchange time distributions at
low temperature.  In order to investigate the structure of exchange
time distributions in detail, we define {\it mobility index},
$m_{i}(t)$, that depends upon whether the lattice site is filled with
an excitation or not and also whether it is in an mobile or immobile
configuration. Specifically,
\begin{equation}
m_{i}(t)=4-2 n_{i}(t)-f_{i}(t),
\end{equation} 
where $f_{i}(t)=1$ when the lattice site $i$ is in a mobile
configuration at time $t$, and it is zero otherwise. As such,
$m_{i}(t)$ is 1 when site $i$ is excited and mobile, 2 when it is
excited and immobile, 3 when it unexcited and mobile, and 4 when it is
unexcited and immobile.

\begin{figure}[t]
\begin{center}
\epsfig{file=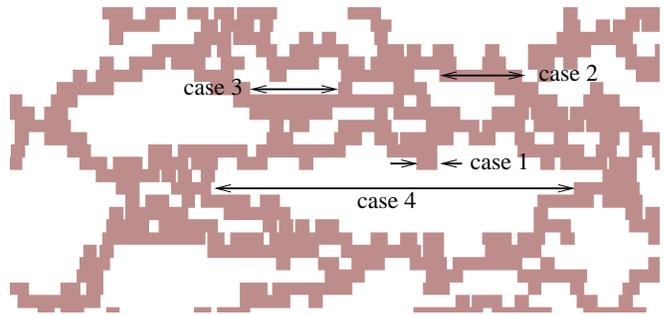, width=\columnwidth} \caption{Examples of exchange
   events that belong to each case of Eq.~(\ref{mobcase}) are
   illustrated in the trajectory of $d=1$ FA model.} \label{traj2}
\end{center}
\end{figure}

During an exchange time of the lattice site $i$, $n_{i}$ does not
change, while its nearest neighbors may make flipping events, thus
changing $m_{i}$ over time.  In order to monitor changes in the local
mobility during an exchange time, it is useful to introduce an
averaged mobility index for a single exchange event,
\begin{equation}
\overline {m}_{i}(t_{\rm s})
= \frac{1}{t_{\rm x}} \int_{0}^{t_{\rm x}} \mathrm{d} t' m_i(t_{\rm s}+t'), 
\label{mbar}
\end{equation} 
where $t_{\rm s}$ is the start time of an exchange event for site $i$,
and the zero time in this formula is the time at which the exchange
event begins. See Fig.~\ref{traj}.  If there has not been any change
in the local configuration during an exchange time, the lattice site
remains always mobile, and the averaged mobility index will be either
$\overline {m}_{i}=1$ (for $n_i=1$) or $\overline{m}_{i}=3$ (for
$n_i=0$). However, when the local mobility changes during the exchange
time period due to changes of the states in nearest neighbors, the
averaged mobility index will be either $1<\overline {m}_{i}<2$ (for
$n_i=1$) or $3<\overline{m}_{i}<4$ (for $n_i=0$).  We define four
different cases of the averaged mobility. Namely,
\begin{equation}
\begin{array}{lccl}
\mbox{case 1} &:& \overline{m}_i =1 & \mbox{($n_i=1$ and $f_i=1$)} \\
\mbox{case 2} &:& 1<\overline{m}_i <2 & \mbox{($n_i=1$ and $f_i$ changes)} \\
\mbox{case 3} &:& \overline{m}_i =3 & \mbox{($n_i=0$ and $f_i=1$)} \\
\mbox{case 4} &:& 3<\overline{m}_i <4 & \mbox{($n_i=0$ and $f_i$ changes)}.
\end{array}
\label{mobcase} \end{equation} 
Examples of exchange times that belong to each case are shown in the
trajectory picture of $d=1$ FA model in Fig.~\ref{traj2}.

\begin{figure}
\begin{center}
\epsfig{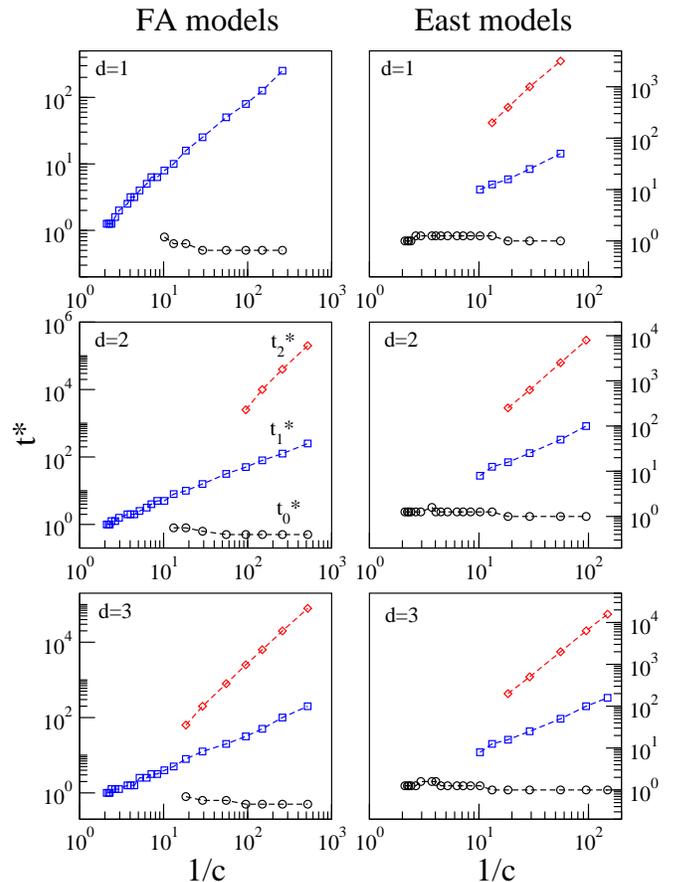}
\caption{Positions of peaks in the exchange time distributions are
shown for various excitation concentrations.}
\label{tpeak}
\end{center}
\end{figure}

To illustrate the correlation between exchange times and environments,
Fig.~\ref{decomp} shows exchange time distributions and their
sub-distributions, each corresponding to one of the cases in
Eq.~(\ref{mobcase}), for $d=1$, 2, and 3 FA and East models at
$T=0.3$.  There exist multiple peaks in the distributions of exchange
times at low temperature in all the cases shown in Fig.~\ref{decomp}.
Positions of peaks in the distribution, $t_{0}^{*}$, $t_{1}^{*}$, and
$t_{2}^{*}$, are shown in the case of $d=3$ FA model.  Temperature
dependence of these peaks for $d=1$, 2, and 3 FA and East models are
presented in Fig.~\ref{tpeak}, and in all the cases, each peak
position is Arrhenius,
\begin{equation} t_{n}^{*}\sim c^{-n} \end{equation} 
where $n=0$, 1, and 2.

To reveal the physical origins of each peak in the exchange time
distributions, the temperature dependence of each peak in the exchange
time distribution is shown in Fig.~\ref{tpeak}.  Cases 1 and 2 will
involve disappearances of excitations, while cases 3 and 4 appearances
of excitations in the trajectory space.  The first peak in the
exchange time distribution corresponds to fast processes of mobile
$n=1$ states embedded in the excitation line. In this case, the
excited state will quickly de-excite in time $t_{\rm 0}^{*} \sim
c^{0}$, and the position of the first peak in the exchange time
distribution does not depend on the temperature as shown in
Fig.~\ref{tpeak}. This process will have the averaged mobility index,
$\overline{m}=1$, corresponding to case 1 in Eq.~(\ref{mobcase}).

The mobile region located at the boundary of excitation line will have
a probability of becoming immobile due to changes in its nearest
neighbors and becoming mobile later. This process corresponds to case
2 with the averaged mobility index $1<\overline{m}<2$. It is
responsible for fluctuations of thicknesses of excitation lines in
trajectory space, and it gives rise to the second peak in the exchange
time distribution, with temperature dependence $t_{\rm 1}^{*} \sim
c^{-1}$.

Dynamical exchange processes corresponding to case 3 also contribute
to the second peak in the distribution. In this case, $n=0$ state next
to an excitation line becomes excited in time $t_{1}^{*}\sim
c^{-1}$. This process involves a creation of an excitation next to a
pre-existing excitation line. In terms of the geometry of the
trajectory space, exchange events corresponding to cases 2 and 3 are
responsible for bending of excitation lines.

Finally, the exchange events that have the longest times, resulting in
the third peak, will correspond to the up-flip event of the $n=0$
states far away from the excitation lines. This case has the averaged
mobility index of case 4. In $d=1$ FA model, for example, it
corresponds to a region deep inside the bubble structure in the
trajectory space.

\section{Distributions of Exchange and Persistence Times}
\label{sec5}

\subsection{Strong Glass Former Models}

\begin{figure}
\begin{center}
\epsfig{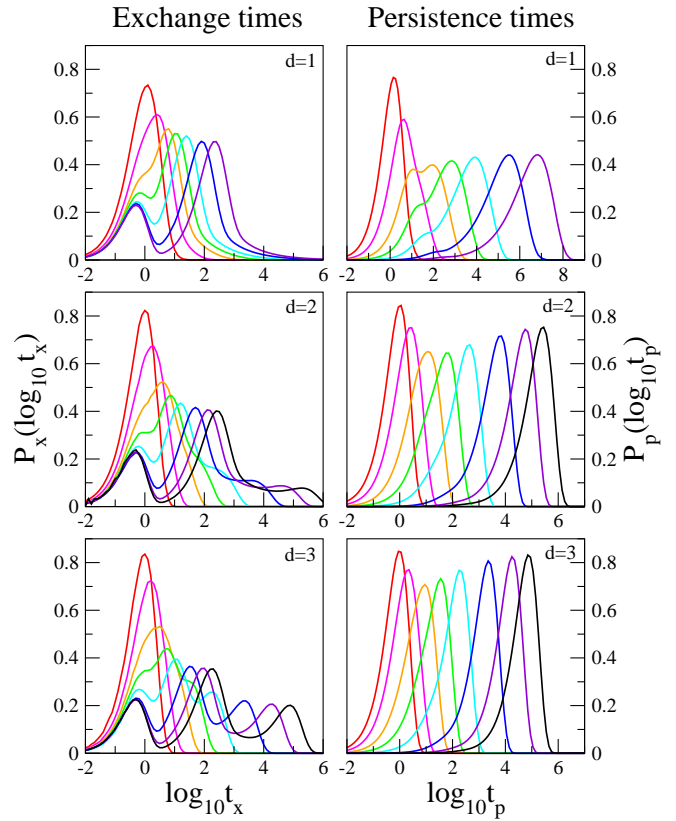}
\caption{Distributions of the logarithm of exchange and persistence
   times of FA model for $d=1$,2, and 3 at various temperatures. From
   left to the right: $T=$ 10, 1, 0.55, 0.4, 0.3, 0.22, 0.18 for $d=1$
   case, and $T=$ 10, 1, 0.55, 0.4, 0.3, 0.22, 0.18, 0.16 for $d=2$
   and $d=3$ cases.
\label{pexppefa}}
\end{center}
\end{figure}

Figure \ref{pexppefa} compares the exchange time distributions (left
panels) and persistence time distributions (right panels) for $d=1$,2,
and 3 FA models at various temperatures. Exchange time distributions
shown in the left panel depend strongly on the temperature. In the
high temperature regime, the exchange time distribution has a single
peak because there is no clear distinction between different exchange
events due to the mean-field nature of the dynamics in the high
temperature regime. However, as temperature decreases, the mean-field
picture is no longer valid, excitation lines dominate trajectory
space, and different kinds of dynamical exchange events manifest
themselves in the distribution as distinct peaks. This fact is evident
from the multiple peak structures in the exchange time distribution in
all dimensions in the FA model. In $d=3$ FA model, for instance,
distinct triplet structure is observed when $T=0.4$ and below. Each
peak corresponding to different dynamical exchange event exhibits
different behaviors of temperature dependence as presented in
Fig.~\ref{tpeak}.

As the dimensionality increases, the exchange time distribution shows
more and more well--separated triplet peak structure in the low
temperature regime. The relative weight of the third peak increases as
the dimensionality increases.

The persistence time distributions shown in the right panel of
Fig.~\ref{pexppefa} are less structured than the exchange time
distributions, as to be expected from Eq.~(\ref{ppepex}).  As the
temperature decreases, the single peak in the persistence time
distribution moves towards the long time region. Physically,
persistence time distributions are statistical measures of duration of
a state at an arbitrary space-time point in the trajectory
space. Thus, relative to exchange time distributions, persistence time
distributions emphasize processes occurring over longest times. The
greater structure in the exchange time distributions, relative to the
persistence time distributions, is due to the former giving equal
weights to events, irrespective of their durations.

\begin{figure}
\begin{center}
\epsfig{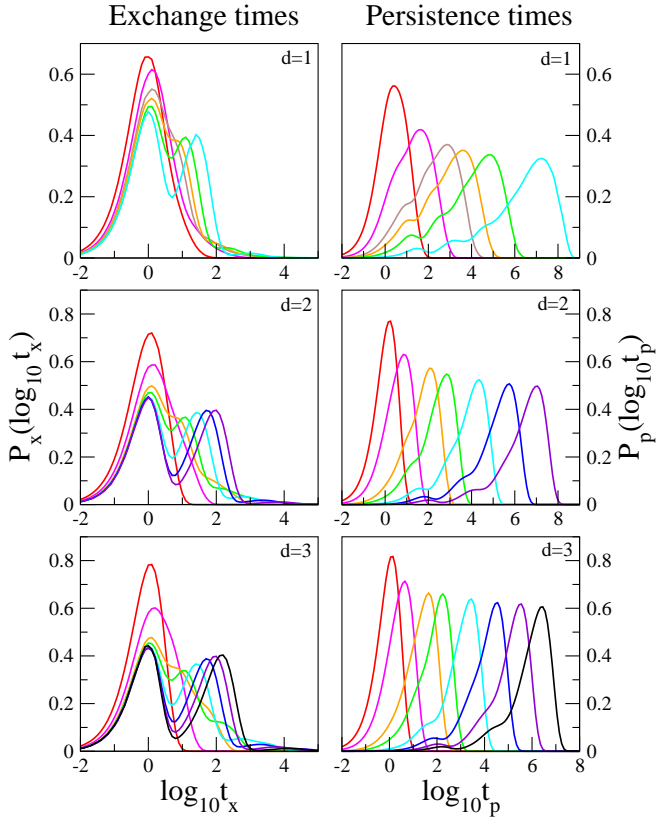}
\caption{Distributions of the logarithm of exchange and persistence
times of East model for $d=1$, 2, and 3 at various temperatures. From
   left to the right: $T=$ 10, 1, 0.6, 0.5, 0.4, and
0.3 for $d=1$ case, $T$=10, 1, 0.5, 0.4, 0.3, 0.25, and 0.22 for 
$d=2$ case, and $T$=10, 1, 0.5,
0.4, 0.3, 0.25, 0.22, and 0.2 for $d=3$ case. \label{pexppeea}}
\end{center}
\end{figure}

\subsection{Fragile Glass Former Models}

\begin{figure}
\begin{center}
\epsfig{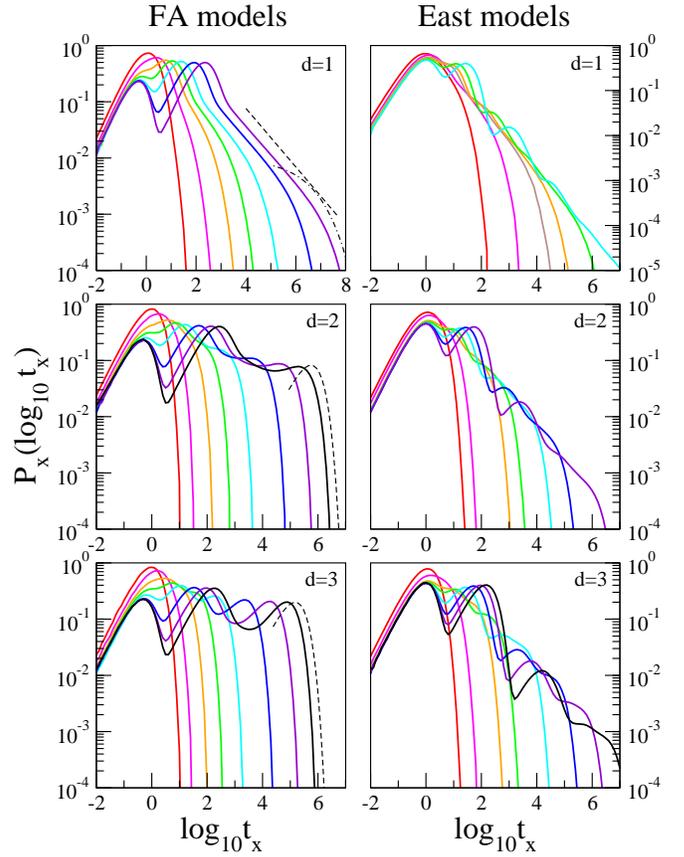}
\caption{Exchange time distributions of FA models (left panels) and
East models (right panels) in log scales. Temperatures are the same as
those given in the captions to Figs.~\ref{pexppefa} and
\ref{pexppeea}. A power-law corresponding to $p_{\rm x}(t_{\rm x})
\sim t_{\rm x}^{-\alpha}$ or $P_{\rm x}(\log_{10} t_{\rm x}) \sim
t_{\rm x} p_{\rm x}(t_{\rm x}) \sim t_{\rm x}^{1-\alpha}$ is shown as
a dashed line with $\alpha =0.52$, and a stretched exponential
corresponding to $p_{\rm x}(t_{\rm x}) \sim e^{-(t_{\rm x})^\beta}$
where $\beta=0.38$ is shown as a dot--dashed line for $d=1$ FA
model. In $d=2$ and 3 FA models, exponential fits, $p_{\rm x}(t_{\rm
x})\sim e^{-t_{\rm x}/\tau}$, are given as dashed lines for the lowest
temperature of each case.} \label{pexfaea}
\end{center}
\end{figure}

Exchange and persistence time distributions for the East model cases
are shown in Fig.~\ref{pexppeea}.  In the high temperature regime, the
exchange time distributions have a single peak as in strong liquids or
FA models. However, in the low temperature regime, in contrast with
its behavior in the FA model, they show two dominant doublet peaks in
all dimensions. We also notice that, albeit small, broad tails develop
in the wings of the distributions at low temperatures. This feature is
especially evident in Fig.~\ref{pexfaea}, where the exchange time
distributions of the FA and East models are compared in log--log
scale. In the case of persistence time distributions shown in the
right panel of Fig.~\ref{pexppeea}, single peaks develop into broad
distributions with shoulders as temperature decreases.

We compare exchange time distributions in FA and East models in
Fig.~\ref{pexfaea}.  In $d=1$ case, exchange time distributions of
both FA and East models show long time tails in the low temperature
regime in common. Additionally, the distributions in the East model
exhibit logarithmic oscillations. In higher dimensions, $d=2$ and 3,
the exchange time distributions in FA models exhibit distinct
multiplet peak structures due to their mean-field natures in higher
dimensions. However, the distributions in the East models remain
similar to those in $d=1$ cases, exhibiting long time tails and
superimposed logarithmic oscillations. These are due to the
hierarchical dynamics that arises from directed facilitation in the
East models at all dimensions.

In the $d=1$ FA case, due to the formations of bubble-like structures
in the trajectory space, the exchange time distributions decay with
slowly decaying long time tails.  It turns out that in $d=1$ FA model
case the exchange time distribution first decays as a power-law
(dashed line) then turns into a stretched exponential (dot--dashed
line) as shown in Fig.~\ref{pexfaea}.  However, in $d=2$ and $d=3$ FA
models, they decay as an exponential at the long time region shown as
dashed lines. This indicates that in higher dimensions the FA models
are less ``glassy'' than $d=1$ case. In higher dimensions, chances of
forming bubble--like structures will be much smaller than in $d=1$
case. In the East model cases, similar behaviors are observed in all
dimensions.  Namely, the exchange time distributions decay with
multiple transitions superimposed with stretched exponentials in the
low temperature regime.

\begin{figure}
\begin{center}
\epsfig{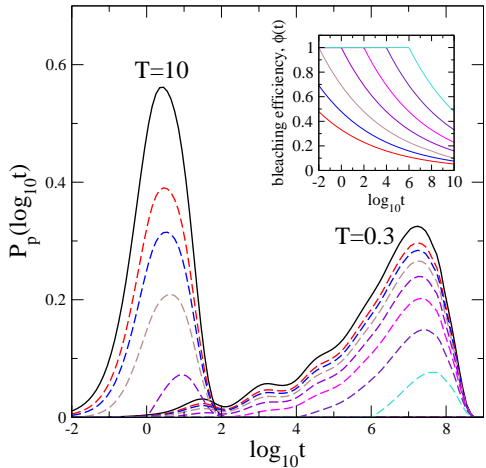}
\caption{Persistence time PDF's at a high ($T=10$) and a low ($T=0.3$)
temperature at equilibrium (solid lines) and right after the bleaching
(dotted lines).  Each dashed line corresponds to the distribution
created by bleaching efficiency function, $\phi(t)$ with different
values of $\tau$, given in the inset of Fig.~(\ref{Ptbl}).  Inset:
Bleaching efficiency function used in the simulation, Eq.~(\ref{bl}),
with $\alpha=0.08$. From left to the right, $\tau$ progressively
increases from $\tau=10^{-6}$ to $\tau=10^6$ by a factor of 100.}
\label{Ptbl}
\end{center}
\end{figure}

\section{Simulations of Bleaching Experiments}
\label{sec6}

Experiments have been performed in order to measure the timescale of
the approach of non--equilibrium distributions of local relaxation
times to equilibrium. 
\cite{ediger-arpc-00,wang-jpcb-99,schmidt-rohr-prl-91,bohmer-epl-96}
The timescale at which the non--equilibrium distribution approaches to
equilibrium will be called the {\it recovery time}.  Based on the $d=1$
East model, we present results of numerical simulations of one of such
experiments - dynamical hole burning experiments or bleaching
experiments. \cite{ediger-arpc-00,wang-jpcb-99}

\begin{figure}
\begin{center}
\epsfig{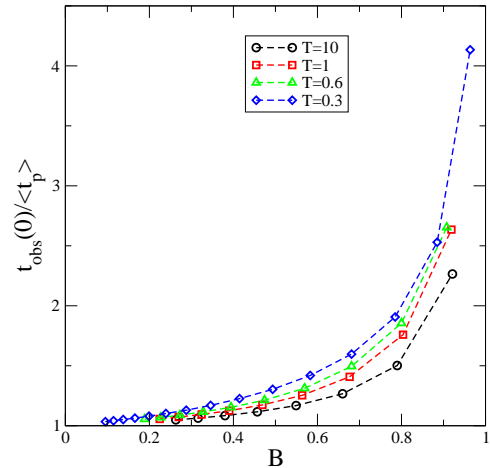}
\caption
{ Mean persistence times right after the bleaching experiment, $t_{\rm
obs}(0)$, are shown for various bleaching depths and
temperatures. $t_{\rm obs}(0)$ increases more abruptly as a function
of the bleach depth at a lower temperature.  }
\label{figtobs}
\end{center}
\end{figure}

In numerical simulations of the bleaching experiment, the bleaching
efficiency function, $\phi(t)$, is introduced as a probability of
bleaching out the region that has a local persistence time of $t$. We
choose the following form as $\phi(t)$, 
\begin{equation} 
\phi(t) = {\rm min} \left[1, (\tau/t)^\alpha\right],\label{bl} 
\end{equation} 
where $\tau$ and $\alpha$ can be varied. We choose $\tau=1$, and $\alpha=0.08$
motivated by experimental results in Ref.~\onlinecite{wang-jpcb-99}.  After a
lattice site is chosen to be ``bleached out'' in simulations,
persistence times of that site are not collected at later times as
time goes on.

Numerical values of the bleaching efficiency functions are shown in
the inset of Fig.~\ref{Ptbl}. The bleaching depth, $B$, is defined as
the fraction of lattice sites that are bleached out by applying the
bleaching efficiency function, and is given by 
\begin{equation} 
B = \int_{0}^{\infty} \mathrm{d} t \phi(t) p_{\rm p}(t). \label{B} 
\end{equation} 
Probability distributions of persistence times after the bleaching are
shown in Fig.~\ref{Ptbl} for two different values of the temperature
and various values of $\tau$.

Immediately after the bleaching, the system is out of equilibrium,
with a non-equilibrium persistence time distribution given (to within
a normalization constant) by
\begin{equation} 
p_{\rm ne}(t) = \frac{p_{\rm p}(t) (1-\phi(t)) }{{1-B}}.
\end{equation} 
After waiting a time $t_{\rm w}$, this distribution relaxes to
$p(t;t_{\rm w})$, where $p(t;0)$ is $p_{\rm ne}(t)$ and $p(t;\infty)$
is the equilibrium $p_{\rm p}(t)$.  $t_{\rm obs}(t_{\rm w})$ is the
first moment of $p(t;t_{\rm w})$,
\begin{equation} 
t_{\rm obs}(t_{\rm w}) = \int_{0}^{\infty} \mathrm{d} t \ t \ p(t;t_{\rm
w}), \label{tobs} 
\end{equation} 
and $t_{\rm obs}(0) > \langle t_p \rangle$.  $t_{\rm obs}(0)$ at
various bleaching depths and temperatures are shown in
Fig.~\ref{figtobs}.

\begin{figure}
\begin{center}
\epsfig{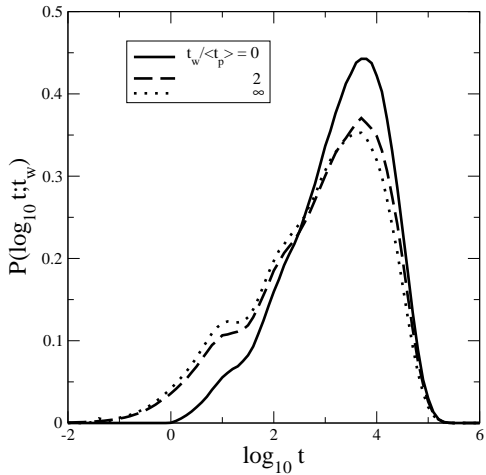}
\caption
{ Relaxation dynamics of the distributions of persistence times to
equilibrium after the bleaching of $d=1$ East model at
$T=0.5$. $\tau=1$ and $\alpha =0.08$ were chosen in $\phi(t)$.
}\label{Pt}
\end{center}
\end{figure}

The relaxation dynamics is monitored by following a time evolution of
the non--equilibrium persistence time distribution obtained from
un-bleached lattice sites for various waiting times, $t_{\rm w}$.
Figure \ref{Pt} shows the relaxation dynamics of the initial
non--equilibrium distribution of persistence times for $d=1$ East model
at a low temperature.

\begin{figure}
\begin{center}
\epsfig{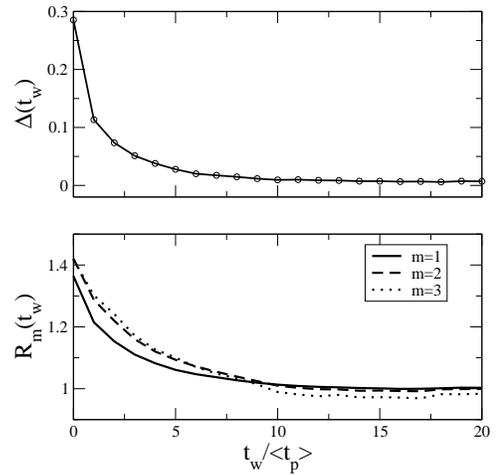}
\caption
{ Recovery of the non--equilibrium distribution, $p(t;t_{\rm w})$ to
equilibrium is shown by the decay of $\Delta(t_{\rm w})$ (upper panel)
and $R_m (t_{\rm w})\equiv \langle (t_{\rm obs}(t_{\rm w}))^m \rangle
/ \langle (t_{\rm p})^m\rangle$ (lower panel).  Timescale of the
recovery to equilibrium is on the order of the equilibrium mean
persistence time itself. The model parameters are the same as those
used in Fig.~\ref{Pt}.  } \label{ptevol}
\end{center}
\end{figure}

The timescale of the relaxation process of the non--equilibrium
distribution can be estimated by several means.  For example, one can
look at the waiting time dependence of moments of the
distributions. Also, it is possible to measure how far a
non--equilibrium distribution is separated from equilibrium by
defining a distance between two distributions by
\begin{equation}
\Delta(t_{\rm w}) \equiv \int_{0}^{\infty} \mathrm{d} t \left| p(t;t_{\rm
w}) - p_{\rm p}(t) \right |. \label{dist}
\end{equation}
$\Delta(t_{\rm w})$ and the first three moments of the
non--equilibrium distribution are shown in Fig.~\ref{ptevol} for
different values of the waiting time. From these calculations, the
timescale of a recovery to equilibrium is estimated to be on the order
of the equilibrium mean persistence time, $\langle t_p\rangle$. This
predication is at variance with results from bleaching experiments, 
 \cite{ediger-arpc-00,wang-jpcb-99} but is is consistent with results
from NMR experiments. \cite{schmidt-rohr-prl-91,bohmer-epl-96,qian-epjb-00} The
reason(s) for this discrepancy remains to be clarified.

\section{Discussion} \label{sec7}

In this paper, we have provided detailed illustrations of the 
generally large differences between distributions of persistence 
times and exchange times.  The distributions are the same only when 
these distributions are Poissonian.  Thus, the differences are 
manifestations of dynamic correlations, correlations that exist due 
to dynamic heterogeneity.  Despite the differences between the two 
distributions, the behavior of one determines the other, as
Eq.~(\ref{ppepex}) shows.

The first moment of the exchange time distribution is Arrhenius, 
independent of model and dimensionality.  This result can be 
important if an experimental means is available to measure this 
moment.  In particular, Eq.~21 shows that the measurement of the mean 
exchange time determines the average mobility concentration, $c$, the 
central control parameter in our description of glassy dynamics.  
 \cite{garrahan-prl-02,garrahan-pnas-03,berthier-pre-03,jung-pre-04,berthier-epl-05}
This can be accomplished with single-molecule experiments in which 
exchange times for a probe molecule passing between fast and slow 
environments are measured directly. \cite{vandenbout} 
 The determination of $c$ in principle enables 
parameter-free tests of scaling relations we have predicted for 
transport properties.

Significant experimental manifestations of the differences between 
exchange-time distributions and persistence-time distributions are 
decoupling of different measures of dynamics.  The breakdown in 
Stokes--Einstein relations discussed in this paper is one example. 
Another is the existence of the Fickian crossover, marking the 
coarse-graining length-scale beyond which random-walk diffusion is 
correct. \cite{berthier-epl-05}  
At large wave-vectors, dynamics is dominated by 
persistence and molecular motion is not Fickian.  At small 
wave-vectors, dynamics is dominated by exchange, and molecular motion is Fickian.

Thus, there are many different relaxation times for slow dynamics in 
a glass former.  Those connected to the mean exchange time will be 
Arrhenius, even for fragile glass formers.  Non-Arrhenius behavior 
(i.e., fragility), when it exists, is described by the second and 
higher moments of the exchange-time distribution.  This feature of 
dynamics in glass formers is general, not a consequence of a 
particular model.  It follows from the presence of dynamic 
heterogeneity and its consequential fluctuation dominance of
dynamics.

\section*{Acknowledgment}
We are grateful to M.D. Ediger, D.A. Vanden Bout, and A. Heuer for
useful discussions.  This work was supported at Berkeley by the Miller
Research Institute for Basic Research in Sciences (YJ) and by the US
Department of Energy Grant No.\ DE-FG03-87ER13793 (DC), and at
Nottingham by EPSRC grants no.\ GR/R83712/01 and GR/S54074/01 and
University of Nottingham grant no.\ FEF 3024 (JPG).

\newpage

\end{document}